\documentclass{aa}  

\usepackage{graphicx}
\usepackage{txfonts}
\usepackage{physics}
\usepackage{color}
\usepackage{multirow}
\usepackage{tablefootnote}
\usepackage{mathtools}
\usepackage{microtype}
\usepackage{amsmath}
\usepackage{natbib}
\usepackage{adjustbox}
\usepackage[normalem]{ulem}
\usepackage{bm}

\begin{document}

   \title{XMM-Newton observations of the TeV-discovered\\ supernova remnant HESS J1534-571}

   \author{N. T. Nguyen-Dang
          \inst{1},
          G. Pühlhofer
          \inst{1},
          M. Sasaki
          \inst{2},
          A. Bamba
          \inst{3,4,5},
          V. Doroshenko
          \inst{1},
          \and
          A. Santangelo\inst{1}
          }

   \institute{Institut für Astronomie und Astrophysik Tübingen (IAAT), Sand 1, 72076 Tübingen, Germany   \\            \email{dang-thanh-nhan.nguyen@astro.uni-tuebingen.de}
         \and Dr. Karl Remeis Observatory, Erlangen Centre for Astroparticle Physics, Friedrich-Alexander-Universit\"{a}t Erlangen-N\"{u}rnberg, Sternwartstra{\ss}e 7, 96049 Bamberg, Germany
         \and Department of Physics, Graduate School of Science, The University
of Tokyo, 7-3-1 Hongo, Bunkyo-ku, Tokyo 113-0033, Japan
         \and Research Center for the Early Universe, Graduate School of Science, The University of Tokyo, 7-3-1 Hongo, Bunkyo-ku, Tokyo 113-0033, Japan
         \and Trans-Scale Quantum Science Institute, The University of Tokyo, Tokyo  113-0033, Japan
             }

   \date{Received May 24, 2023; accepted July 14, 2023}

\abstract{
We report the results obtained from \textit{XMM-Newton} observations of the TeV-detected supernova remnant (SNR) HESS J1534-571.
We focus on the nature of the cosmic-ray particle content in the SNR, which is revealed by its $\gamma$-ray emission.
No signatures of X-ray synchrotron emission were detected from the SNR. This is consistent with earlier results obtained with \textit{Suzaku} from other regions of the object. A joint modeling of the \textit{XMM-Newton} and \textit{Suzaku} spectra yields an upper limit for the total X-ray flux from the SNR area of $\sim$ 5.62$ \times 10^{-13} \ \mathrm{erg\ cm^{-2}\ s^{-1}}$   (95\% c.l.) in the energy band of 2-10\,keV, for an assumed photon index of 2.0. On the other hand, we do find evidence in the \textit{XMM-Newton} data
for a line-like emission feature at 6.4\,keV from localized regions, again confirming earlier \textit{Suzaku} measurements. We discuss the findings in the context of the origin of the observed $\gamma$-ray emission. Although neither hadronic nor leptonic scenarios can be fully ruled out, the observed line emission can be interpreted as the result of interactions between lower energy ($\sim$ MeV) cosmic-ray protons with high gas density regions in and
around HESS J1534-571, and thus potentially be associated with particles accelerated in the SNR.
\\

}

\titlerunning{XMM-Newton observations of the TeV-discovered supernova remnant HESS J1534-571}
\authorrunning{N. T. Nguyen-Dang}
\maketitle

\section{Introduction}
\label{sec:intro}

Supernova remnants (SNRs) are generally considered as the most relevant sources of energy input for Galactic cosmic rays (CRs) \citep{1964ocr..book.....G,2005JPhG...31R..95H}. The main arguments are the required kinetic energy per unit time, and the fact that diffusive shock acceleration \citep{2001MNRAS.321..433B, 2004MNRAS.353..550B} is a well-established particle acceleration mechanism, in agreement with many observational parameters \citep{1987PhR...154....1B,2001RPPh...64..429M, 1991SSRv...58..259J}. A quantitative probe of this scenario is the electromagnetic emission from the accelerated particles in and around the sources. To specifically probe the high end of the leptonic particle spectra, X-ray synchrotron and TeV $\gamma$-ray Inverse Compton emission are the relevant channels \citep{1998ApJ...493..375R}. The hadronic component, which is energetically relevant from the CR standpoint, can only be probed in $\gamma$-rays (through the $\pi^0$ -decay process), with the high end of the spectrum at TeV-PeV photon energies \citep{1997ApJ...490..619S, 1994A&A...287..959D}.\\

Young SNRs with ongoing particle acceleration towards the highest energies are key targets for the study of CR acceleration, specifically since the escape of particles upstream of the forward shock plays a strong role in modifying the particle spectrum soon after the shock is slowing down. However, the typically high synchrotron fluxes and the lack of spectra extending to $\sim$100 TeV have motivated preference for a leptonic interpretation of the $\gamma$-ray spectra in several cases, despite the need to invoke low B-fields in the bulk of the emission regions (well below $\sim$100 $\mu$G levels needed for efficient highest-energy particle acceleration) \citep{2005JPhG...31R..95H}. TeV-selected SNRs, which should still be (moderately) young, given their TeV dominance and by definition presenting low levels of synchrotron emission, are therefore important targets in view of a potential hadronic dominance of the $\gamma$-ray emission, which would permit a direct view on the hadronic energetics of the objects.\\

This motivates a detailed study of HESS J1534-571 as presented in this paper. HESS J1534-571 is one of only three objects that were classified as SNR candidates after their discovery in TeV $\gamma$-rays, based on a morphological characterization of new objects discovered in the H.E.S.S.\ Galactic plane survey \citep{2018A&A...612A...8H}. From the subsequent identification with a SNR candidate SNR G323.7–1.0 found in archival radio continuum data from the Molonglo Galactic Plane Survey (MGPS, \cite{green2014second}), HESS J1534-571 was then classified as confirmed SNR. No clear X-ray counterpart of the SNR shell has been found so far, with the best limits for the nonthermal X-ray emission obtained with \textit{Suzaku} observations of a fraction of the shell, yielding $2.4 \times 10^{-11}\ \mathrm{erg\ cm^{-2}\ s^{-1}}$ in the energy band 2-12\,keV assuming a power-law model of index 2 and scaling to the full SNR area \citep{2018A&A...612A...8H}. The object has a diameter of $\sim 0.8^{\circ}$ and requires several pointings of current (e.g., \textit{XMM-Newton}) or recent X-ray (e.g., \textit{Suzaku}) satellites for full coverage. Here, we present results obtained from a $\sim$25\,ks \textit{XMM-Newton} pointing covering the previously unobserved part of the SNR. Similar to the \textit{Suzaku} observations that had covered the other parts of the shell, no evidence for X-ray synchrotron emission is found. Using both data sets together, we obtain for the first time robust constraints on the X-ray emission from the entire shell, and we discuss the results in the context of the origin of the observed $\gamma$-ray emission from the object.

Besides synchrotron emission from relativistic electrons and plasma emission (bremsstrahlung and thermally excited fluorescence emission lines), also cold gas can give rise to fluorescence lines in X-ray spectra of SNRs. This process occurs when low energy cosmic ray protons interact with nearby cool gas, where the collisional excitation is then followed by radiative de-excitation. In particular, the fluorescence yield of neutral ion (Fe I) for a de-excitation to K-shell transition is 34\%, much larger than other atoms such as neutral silicon at 5\% or neutral oxygen at 0.8\% \citep{1979JPCRD...8..307K,kallman2004photoionization,2012A&ARv..20...49V}. Due to the high fluorescence yield and the relatively high abundance, the Fe K$\alpha$ line at 6.4\,keV is an important diagnostic channel to study SNRs and the properties of nearby molecular clouds. Evidence for this line has been widely found for different SNRs, such as W28, Kes67, Kes69, Kes78 and W44 \citep{2018ApJ...854...87N}. \cite{saji2018discovery} has reported the existence of such a line in the \textit{Suzaku} observations of HESS J1534-571 at a $4\,\sigma$ significance level. We have extended the investigation of this line to the \textit{XMM-Newton} data presented in this work.

A search for a possible gas association of the source conducted by \cite{maxted2018searching} using Mopra CO Galactic Plane Survey data \citep{2013PASA...30...44B} and the Southern Galactic Plane Survey HI data \citep{2005ApJS..158..178M} indicated that the SNR is possibly located inside the Scutum-Crux arm and could stem from a core-collapse event at a kinematic distance of $\sim$ 3.5 kpc. In this paper, we adopt this distance estimate.\\

The paper is organized as follows: In section \ref{sec:Observations}, the \textit{XMM-Newton} observations and the data reduction is described. In section \ref{sec:Results} we describe the obtained results and put them into context of broadband spectral models describing the emission from assumed non-thermal particle populations. The results are further discussed in section \ref{sec:Discussion} and are put into context of results of other SNRs. We conclude with a summary in section \ref{sec:Conclusion}.

\section{Observations and data reduction
\label{sec:Observations}}

The first attempt to observe HESS J1534-571 with \textit{XMM-Newton} \citep{2012arXiv1202.1651L} was cancelled due to the outburst of a transient source in the field of view. Close to the geometrical center of HESS J1534-571, a previously unknown X-ray binary (XRB) was discovered with the MAXI/GSC instrument in 2017 \citep{2017ATel10699....1N} and independently with \textit{Swift}/BAT \citep{2017ATel11007....1K}, named MAXI J1535-571. Given the uncertain distances, an association between the two objects can neither be claimed nor be disputed at the moment. Observationally, when the XRB is in outburst, \textit{XMM-Newton} observations of the SNR shell are impossible because the wings of the point spread function (psf) pollute the entire European Photon Imaging Camera (EPIC) field of view (FoV). Therefore, observations that had been approved for 2017 had to be called off, and we were only able to obtain observations in 2020. Parameters of the \textit{XMM-Newton} observation of HESS J1534-571 are shown in Table \ref{tab:table1}.

For data reduction, we use the \textit{XMM SAS} software version 20201028\_0905-19.0.0. In particular, the work package \textit{XMM-Newton} Extended Source Analysis Software (XMM-ESAS) is employed. In order to obtain images and spectra of the source, we follow the standard steps as recommended in the XMM-ESAS cookbook\footnote{https://heasarc.gsfc.nasa.gov/docs/xmm/esas/cookbook/xmm-esas.html}.

\begin{table}
\caption{XMM-Newton and Suzaku data used for the analysis}
    $$
    \begin{array}{p{0.17\linewidth}  p{0.23\linewidth} p{0.15\linewidth} p{0.2\linewidth}}
        \hline
        \hline
            Observation ID & Dates & Exposure time (ks) & Instruments  \\
            \hline
            \multirow{3}{*}{0841440101}& \multirow{3}{*}{2020-Mar-03} & 24.07 & MOS1   \\ 
            & & 24.51   &     MOS2 \\           
            & & 17.07   &     PN \\
            508013010 & 2013-Sep-08 & 36.9  &    XIS0-3\\
            508014010 & 2013-Sep-09 & 36.9  &    XIS0-3\\
            508015010 & 2013-Sep-09 & 36.9  &    XIS0-3\\
            508016010 & 2013-Sep-10 & 36.9  &    XIS0-3\\
            \hline
    \end{array}
    $$
    \label{tab:table1}
\end{table}

\begin{figure}
    \centering
    \includegraphics[width=3in]{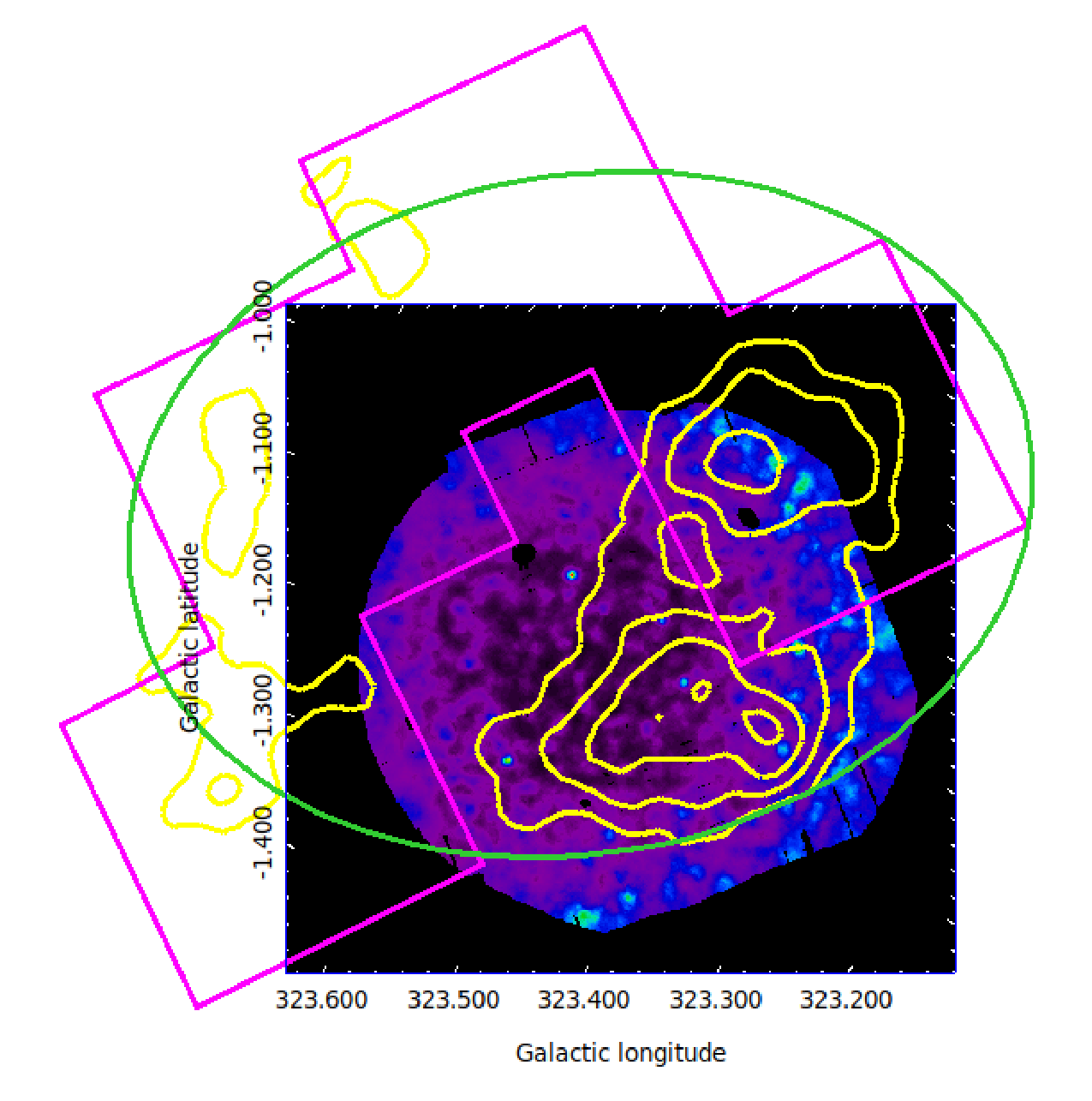}
    \caption{Adaptively smoothed \textit{XMM-Newton} image in the energy band 2.0-5.0\,keV of HESS J1534-571. Green ellipse shows the radio boundary, magenta region indicates 4 \textit{Suzaku} poitings and yellow lines represent the TeV surface brightness contour.}
    \label{fig:xmm-fov}
\end{figure}
First, the standard re-processing was applied using the ESAS task \textit{epchain, emchain}. The events are then filtered for soft proton flares, at the same time assorted diagnostic files are created using \textit{pn-filter, mos-filter} with default criteria. The MOS CCDs were operated in a full frame mode. After examining the CCDs for their anomalous states, we exclude CCD \#3 and CCD \#6 of EPIC MOS1 from the analysis. MOS1 CCD \#6 was lost due to the micrometeorite strike in March 2005 and MOS1 CCD \#3 was damaged in December 2012. Point sources are listed for later removal from the entire FoV using the task \textit{cheese}, applying a flux threshold at $1\times 10^{-14}\ \mathrm{ergs\ cm^{-2}\ s^{-1}} $.

Consequently, spectra, response files, and the model particle background spectra and images are produced with the use of the FWC data via the tasks \textit{pn-spectra, mos-spectra}. Finally, tasks \textit{pn\_back, mos\_back} are run to create quiescent particle background files. The pipeline makes use of the data from unexposed corners of the cameras, filter wheel closed data as well as the \textit{ROSAT} All-Sky Survey in order to estimate the contribution of cosmic rays (\cite{kuntz2008epic}). The spectra are then grouped by the HEASOFT FTOOLS command \textit{grppha}, with minimum 50 counts per bin. The coverage areas of each instrument are calculated by the task \textit{proton\_scale}. The spectral analysis is carried out with \textit{xspec} software version 12.11.1b.

The \textit{XMM-Newton} images presented in this work are vignetting corrected, background subtracted and point-source removed. After the initial screening and filtering, residuals from the soft proton contamination might still be present in the images and spectra. In order to take these soft proton residual into account, in addition to the model that represent the main spectra, we added a power law model that is not folded with the instrumental effective areas. The best fit parameters for the soft proton contamination are used to create soft proton images and to be subtracted from the final images. For each energy band, the images from individual instruments are combined and adaptively smoothed.

For the purpose of producing an image in a narrow band around 6.4\,keV covering the entire source, we have also reprocessed the archival \textit{Suzaku} data taken on HESS J1534-571. \textit{Suzaku} was pointed towards the northern and eastern part of the source in four dedicated observations of approximately 40 ksec each in 2013 \citep{saji2018discovery}. We employ the HEASOFT software version 6.24, \textit{Suzaku} reprocessing version 3.0.22.44, and the remote calibration files to make sure that we have access to the most current calibration tools. We filter bad pixels, exclude the data from the chip corners, combine 3x3 and 5x5 modes where possible, exclude data with elevation < 5 deg above Earth or < 20 deg above sunlit limb of Earth. Extra event filtering was applied for XIS0 due to the area discrimination after the  XIS0 anomaly in 2009. For spectral analysis, the ancillary response files (ARF) and Redistribution matrix files (RMF) are created using the tasks \textit{xissimarfgen} and \textit{xisrmfgen}, respectively. For image production, the non-X-ray background within 300 days from the observation is derived by the task \textit{nxbgen} for background subtraction. A simulated exposure map is also created via the task \textit{xissim} for vignetting correction. The individual images from XIS0, XIS1, XIS3 are rebinned with 8 x 8 pixels and combined using \textit{ximage}. For images above 5\,keV, only the front illuminated CCDs XIS0 and XIS3 are used because the signal-to-noise of the back illuminated XIS1 is worse at this energy range. 
Spectra were extracted from the dashed white regions in Fig. \ref{fig:6365-image-with-suzaku}. These are the enhanced regions in the 6.3-6.5 keV narrow band image, found both by \cite{saji2018discovery} and in this work. The NXB spectra was calculated using data within $\pm 300$ days from the observation day. Similar to \textit{XMM-Newton}, the \textit{Suzaku} spectral analysis is done using the software \textit{xspec}.

For reference, Fig. \ref{fig:xmm-fov} shows the TeV contours of HESS J1534-571 \citep{2018A&A...612A...8H} and the outer boundary of the radio shell G323.7-1.0 discovered in MGPS data \citep{green2014second} as well as the EPIC and X-ray Imaging Spectrometer (XIS) FoVs of the \textit{XMM-Newton} pointing and the archival \textit{Suzaku} pointings \citep{saji2018discovery}, respectively. 

\section{Results}
\label{sec:Results}

\subsection{X-Ray continuum emission}
\label{sec:Results1}
To investigate the potential continuum emission from HESS J1534-571, we produced co-added images from all three \textit{XMM-Newton}-EPIC cameras in two energy bands (the same as used in \citealt{saji2018discovery}), namely a soft band at 0.5-3.0\,keV and a hard band at 5.0-8.0\,keV. The former is sensitive to putative thermal emission with typical SNR temperatures, the latter suited to search for non-thermal (synchrotron) emission. Taking the radio SNR
boundary as a guideline, roughly 85$\%$ of the EPIC FoV covers the SNR, whereas the rest is outside the SNR and can serve as reference and background region. Close inspection of the \textit{XMM-Newton} images shows that there is no obvious structure visible that is associated with the SNR, neither in thermal nor in non-thermal emission.
\\

To be able to compare to spectral models of the broadband
non-thermal emission of the SNR (see Sect. 3.3), we derived a flux upper limit in the 2.0-10.0\,keV band. As on-source region, we used the part of the SNR as defined by the radio boundary that is covered by the EPIC FoV (see Fig. \ref{fig:xmm-fov}). A background spectrum is derived from a source-free (i.e. outside the radio SNR boundary) patch of the FoV. The CCD chips of MOS1 that were available at the time of the observations do not cover the background control region. Therefore, we used the off-source spectrum of MOS2 as the background spectrum for EPIC MOS1, given that the sensitivity and the calibration of the two MOS cameras are similar, resulting systematic errors are minor. \\

\begin{figure}
    \centering
    \includegraphics[width=3.5in]{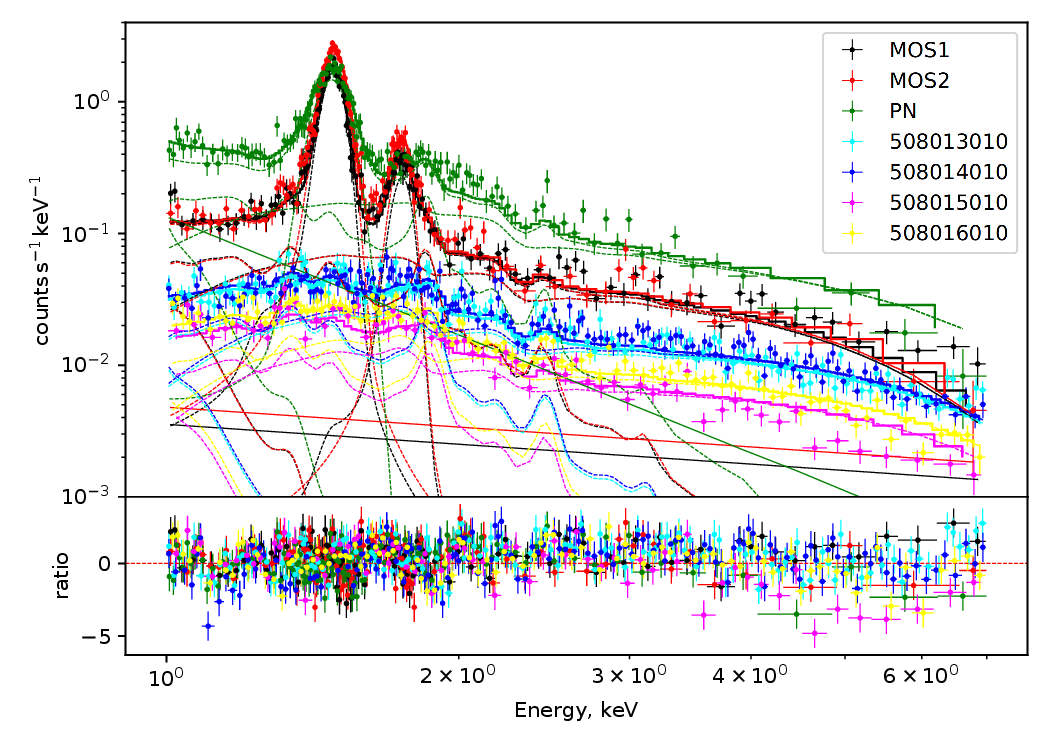}
    \caption{X-ray spectra of the \textit{XMM-Newton} pointing (MOS1, MOS2, PN) and the four \textit{Suzaku} pointings in the energy range 1.0-7.0 keV. The upper panel depicts the data and best-fit total model spectrum as well as the individual model components, whilst the lower panel shows the ratio between each spectrum and the corresponding best-fit model.}
    \label{fig:continuum-spectra}
\end{figure}

We also employ a simultaneous spectral fitting for data from \textit{XMM-Newton} and \textit{Suzaku}. First, test fits for each individual spectrum from each data set (MOS1, MOS2, PN, and 4 \textit{Suzaku} observations) and each telescope (\textit{XMM-Newton} and \textit{Suzaku}) were performed in order to estimate the reasonable range of fit parameters. Then, we carried out a joint fitting to all the source and background data. The background spectra are extracted from the region outside of the radio boundary. We fit all the spectra to the model which consists of the astrophysical background and for \textit{XMM-Newton} data an additional, not ARF-folded model for any residual soft proton contamination. We choose the energy range of 1.0-7.0\,keV because above 7\,keV the count rate is low. The model for the simultaneous fit is then a collisionally-ionized diffuse gas \textit{apec} at plasma temperature 0.1\,keV to represent the local hot Bubble, an absorbed power law \textit{TBabs*pow} of index 1.46 for the unresolved cosmic X-ray background, and another absorbed \textit{TBabs*apec} at 0.65\,keV for the interstellar medium (ISM). The temperature of the ISM is obtained from earlier test fits and then fixed. After the best fit is achieved, we fixed the best fit values and add an absorbed power law \textit{TBabs*pow} model to represent the undetected X-ray source and its total absorption. Given that the relevant absorption to the object's distance is not known, we adopt the column density through the entire Galaxy estimated by the HEASoft NH FTOOL \citep{2016A&A...594A.116H} for the assumed emission component from the SNR. The derived upper limit is therefore a conservative estimate, but the impact is anyway small given the fitted energy range above 2\,keV. Data and best-fit models are plotted in Fig. \ref{fig:continuum-spectra}. The total models (plotted as solid lines) show the base model without the power law component, which indicates the upper limit of the synchrotron emission. The spectra of the background control regions are not plotted for better visual presentation.

The power law represents the expected spectral shape of the synchrotron emission in the considered energy range. We adopt two power-law indices, a generic 2.0 value and a much softer value of 3.0 which corresponds to the average photon index of the model prediction by \cite{araya2017detection}. Finally we report the 95\% confidence level flux for this additional power law component and extrapolate into the range 2.0-10.0\,keV as the flux upper limit from the covered fraction of the SNR (724.84 $\mathrm{arcmin}^2$ for the \textit{XMM-Newton} FoV and 1214.4 $\mathrm{arcmin}^2$ for \textit{Suzaku} FoV). To be able to compare to model expectations, this limit needs to be extrapolated to the area of the entire SNR defined by the radio region (1521.94 $\mathrm{arcmin}^2$). Assuming that the surface brightness of the SNR is the same (or lower) in the not-covered areas, a plain area scaling can be adopted, i.e. the relative error on the uncertainty of the zero flux level remains the same. Here we simply scale the flux density to the area of the radio boundary. A flux upper limit derived from \textit{XMM-Newton} data only, representing the flux coming from the SNR's fraction covered by the \textit{XMM-Newton} FoV is also shown for reference. These two extraction areas are named "Radio boundary" and "XMM-Newton", respectively, see Table \ref{tab:UL-results}.
 \\

To test the robustness of the result, we also used a narrower region around the peak significance region of the TeV emission for extracting an on-source spectrum, given that the X-ray emission may trace the gamma-ray emission region better than the entire SNR. This extraction area is referred to as "TeV peak". The derived limits are compared in Table \ref{tab:UL-results}. As reference result, we choose the limit obtained for the radio SNR area with an index of 2.0, which yields an upper limit of 5.62$ \times 10^{-13} \ \mathrm{erg\ cm^{-2}\ s^{-1}}$ at 95\% confidence level. To put this number into context, the expected flux in a model by \cite{araya2017detection} is $2.7 \times 10^{-14} \ \mathrm{erg\ cm^{-2}\ s^{-1}}$ in the 2-10\,keV energy band. The expectation critically depends on the adopted energy cutoff of the electron spectrum population, which was constrained in a leptonic model by fitting to the GeV-TeV (and radio) data from the source. Further comparisons to synchrotron and broadband model expectations are discussed in section 3.3.\\

\begin{table}
\caption{Results of upper limit estimation. The flux upper limit in the energy range 2.0-10.0\,keV is reported in units of $\times 10^{-13} \ \mathrm{erg\ cm^{-2}\ s^{-1}}$. The definition of the extraction regions is described in the text.}
\begin{adjustbox}{width=3.5in}
$$ 
     \begin{tabular}{l cc cc cc }
    \hline
            Region &
            \multicolumn{2}{c}{Radio boundary} &
            \multicolumn{2}{c}{TeV peak} & \multicolumn{2}{c}{XMM-Newton}\\
            \hline

     Index & $\Gamma$ = 2  & $\Gamma$ = 3  &  $\Gamma$ = 2  &      $\Gamma$ = 3&$\Gamma$ = 2  & $\Gamma$ = 3\\
    $\mathrm{UL}_{2-10}$  &    5.62  & 4.94 & 1.03  &  1.00 &2.52 & 2.27\\
    $\chi^2_{\mathrm{reduced}}$ & 3421/2930& 3414/2930& 2922/2448&2916/2448 &2823/2251 &2818/2251\\

            \hline
         \end{tabular}
        
$$
\end{adjustbox}
     \label{tab:UL-results}
   \end{table}

\subsection{6.4\,keV line emission}
\label{sec:Results2}

\begin{figure}
    \centering
    \includegraphics[width=3.5in]{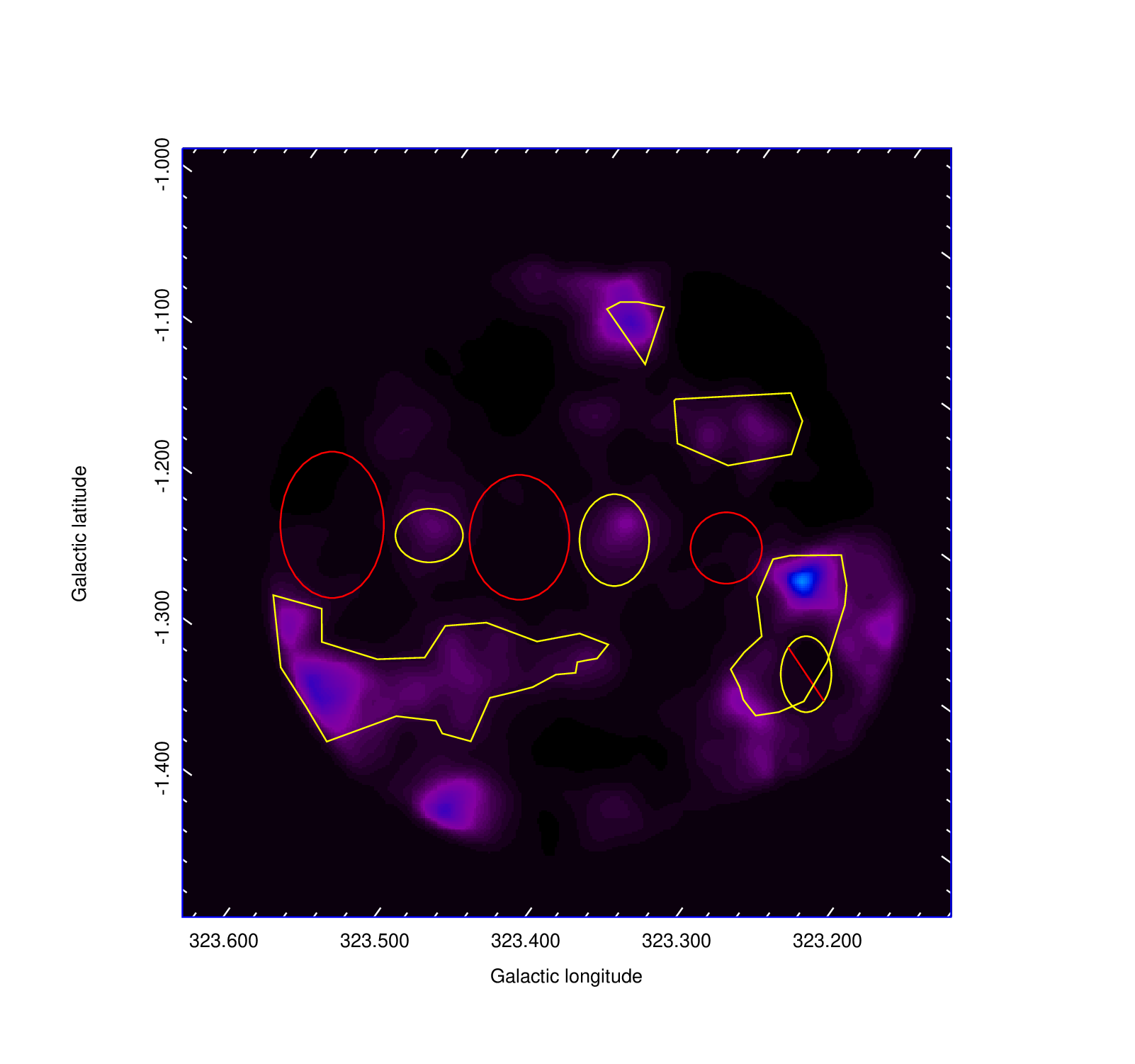}
    \caption{Adaptively smoothed, MOS1-MOS2-PN combined \textit{XMM-Newton} image in the energy band 6.3-6.5\,keV of HESS J1534-571. The enhanced and reference regions of MOS1 are depicted in yellow and red boundaries, respectively. The regions differ slightly for different cameras due to their CCD geometry.}
    \label{fig:6365-regions}
\end{figure}
\begin{figure}
    \centering
    \includegraphics[width=3.5in]{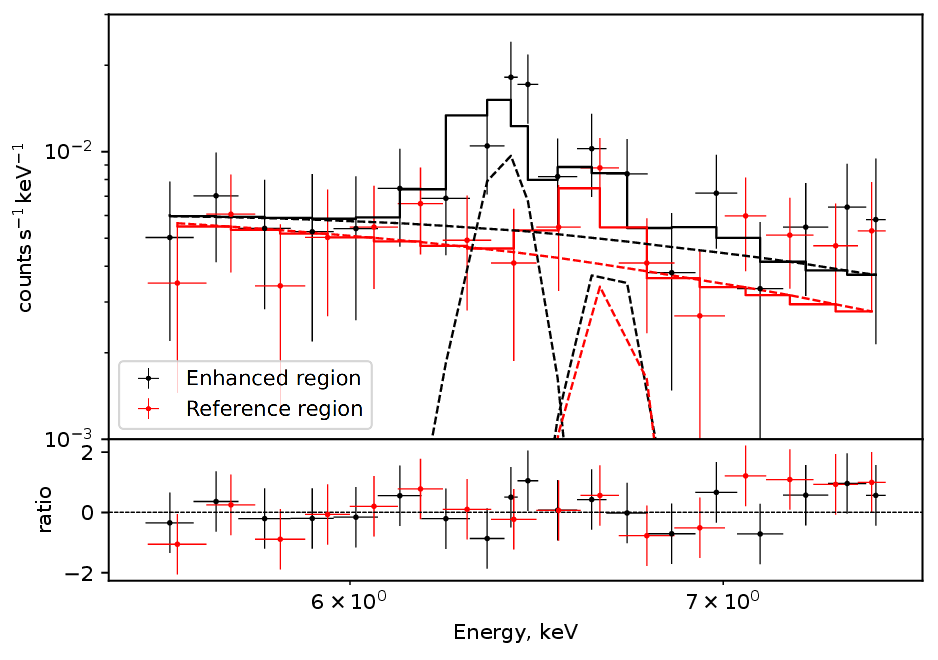}
    \caption{The MOS1+MOS2 stacked spectra and the best fit model of the 6.4\,keV enhanced region (black) and of the reference region (red).}
    \label{fig:6365-spectra}
\end{figure}

Triggered by the report by \cite{saji2018discovery}, we also created an image in the narrow band around 6.3-6.5\,keV. The authors have found evidence for line emission at 6.4\,keV in the four \textit{Suzaku} pointings on HESS J1534-571, at a statistical significance of $\sim4\sigma$. The corresponding narrow-band \textit{XMM-Newton} image also shows evidence for localized emission from the area of the SNR. We chose on-source extraction regions based on the enhancements in the image, and reference background regions. The co-added on- and off-source spectra are shown in Fig. \ref{fig:6365-spectra} in an energy band between 6.3 and 6.5\,keV. \\

Galactic ridge (Galactic Ridge X-ray emission, GRXE) \citep{2016PASJ...68...59Y} is expected to be seen from the direction of HESS J1534-571 because the SNR resides in the Galactic plane. These are unresolved X-ray emission lines along the Galactic plane, namely the neutral iron Fe I K$\alpha$, Fe XXV K$\alpha$, Fe XXVI K$\alpha$, Fe I K$\beta$ centered at 6.4\,keV, 6.68\,keV, 6.97\,keV and 7.06\,keV, respectively. The stacked spectrum from MOS1 and MOS2 data shows that the Fe K$\alpha$ line emission 6.68\,keV from the GRXE is visible in both on-source and off-source spectra at the same level. The Fe XXVI K$\alpha$ line is also detected at 6.97\,keV. The enhancement in the image clearly corresponds to a line-like feature at 6.4\,keV, which can be identified with Fe K$\alpha$ emission, and that is not seen in the reference region. We fit the spectra in the range 5.5-7.5\,keV to a power-law component and four Gaussian lines representing the GRXE (Fig. \ref{fig:6365-spectra}).

\begin{figure}[h!]
\centering
\includegraphics[width=3.5in]{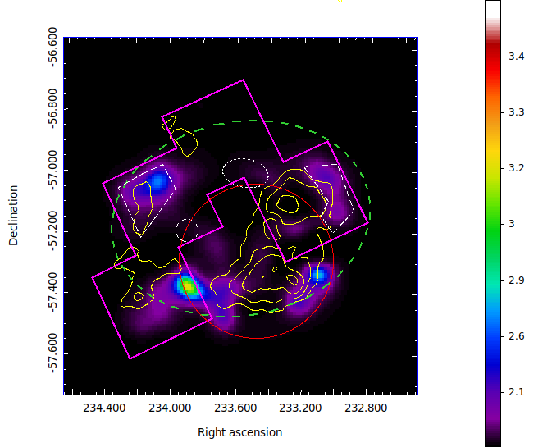}
\caption{Combined \textit{XMM-Newton} and \textit{Suzaku} image of HESS-J1534-571 in 6.3-6.5\,keV, in units of counts per second. The red circle shows the \textit{XMM-Newton} field of view, the green ellipse indicates the \textit{Molonglo} radio boundary, the yellow contour follows the TeV profile, and the magenta region shows the 4 pointings of \textit{Suzaku} observations. The dashed white regions denote the enhancement regions found by \cite{saji2018discovery}. A point source in the Suzaku southern pointing has been removed. Note the similarity between the bright clumps in this narrow energy band and the TeV profile. }
\label{fig:6365-image-with-suzaku}
\end{figure}

Similar to \cite{saji2018discovery}, we fixed the Gaussian width of the line 6.68\,keV at 23 eV \citep{2007PASJ...59S.245K} and linked the normalization of the line 7.06\,keV as 0.125 times that of the 6.4\,keV line \citep{1993A&AS...97..443K}. The best fit parameters are shown in Table \ref{tab:6.4table}. We ran MCMC simulations to calculate the probability that the 6.4\,keV line could stem from background fluctuations. For this, 10.000 spectra in the energy range of interest were simulated. We found that the 6.4\,keV emission is significant at $\sim 3\sigma$ confidence level, and contains the signal from CR-cold gas interaction as well as a small contribution (not significantly detected in the reference region) from the GRXE. Given that the extraction regions had to be chosen based on the low-significance image, this is a pre-trial significance. However, taking the significances from the \textit{Suzaku} and the \textit{XMM-Newton} results together, we consider the results summarized above as a sufficient motivation for further investigation of this feature. Given that the conversion factor from count rates to flux in this band is very similar between the used \textit{XMM-Newton} and \textit{Suzaku} detectors (MOS1+2 and XIS0+3, respectively), we combined the count maps of these two instruments to show the morphology of the enhanced Fe K$\alpha$ line emission across the complete SNR. Fig. \ref{fig:6365-image-with-suzaku} shows a mosaic of \textit{XMM-Newton} and \textit{Suzaku} pointings in the 6.3-6.5\,keV band, with the radio boundary and the TeV emission overlaid as contours. A visual comparison to the \textit{Suzaku} map in \cite{saji2018discovery} reveals overall similar enhancement positions and shapes, despite the limited significances of individual features. Small deviations may be attributed to differences in handling of the area discrimination for the XIS0 chip (lower fluxes in the small dashed circle and ellipse regions in Fig. \ref{fig:6365-image-with-suzaku}) and in the creation of the exposure map. In the southern \textit{Suzaku} pointing, the difference (apparently higher flux) might be attributed to a different zero suppression in the image, after the necessary point source subtraction.

\begin{table}

\caption{Best fit parameters of the spectra analysis of the 6.4\,keV clumps. Errors and upper limits are given at 90\% c.l.}
\begin{minipage}{\textwidth}
\label{tab:6.4table}
    \begin{tabular}{l l l }
    
        \hline
        \hline
         Parameters & Enhanced region & Reference region\\
        \hline
        Energy (keV) & 6.4 (fixed) & 6.4 (fixed) \\
        $\sigma$ (keV) & 0 (fixed) & 0 (fixed) \\
        normalization\footnote{Photon flux in units of $10^{-5} \ \mathrm{photons} \ \mathrm{cm}^{-2}\ \mathrm{s}^{-1}\ $} & $1.892^{+1.017}_{-0.959}$ &  $< 0.679$\\
        Equivalent width\footnote{keV at 95$\%$ level of significance.}  & 549 & 212\\
        \hline
        Energy (keV) & 6.68 (fixed) & 6.68 (fixed) \\
        $\sigma$ (keV) & 0.023 (fixed) & 0.023 (fixed) \\
        normalization & $1.096^{+1.121}_{-0.811}$ &  $0.762^{+0.785}_{-0.628}$\\
        Equivalent width  & 275 & 504\\

        \hline
        Energy (keV) & 6.97 (fixed) & 6.97 (fixed) \\
        $\sigma$ (keV) & 0 (fixed) & 0 (fixed) \\
        normalization & $0.157^{+1.352}_{-0.106}$ & $< 0.578$\\
        Equivalent width & 224 & 172\\

    \end{tabular}
\end{minipage}
\end{table}

\subsection{Broadband SED of HESS J1534-571}
\label{sec:SED}
\begin{figure}
    \centering
    \includegraphics[width=3in]{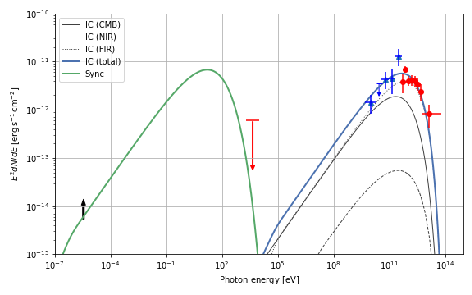}
    \caption{The SED of HESS J1534-571. Plotted as blue triangles are \textit{Fermi}-LAT data \citep{araya2017detection}. Red circles are H.E.S.S. TeV flux points \citep{2018A&A...612A...8H}. The black arrow shows the lower limit of the flux detected in radio \citep{green2014second}. The red arrow illustrates the estimated upper limit in the 2.0-10.0\,keV band. The leptonic model described in the text is also plotted. }
    \label{fig:SED1}
\end{figure}

\begin{figure}
    \centering
    \includegraphics[width=3in]{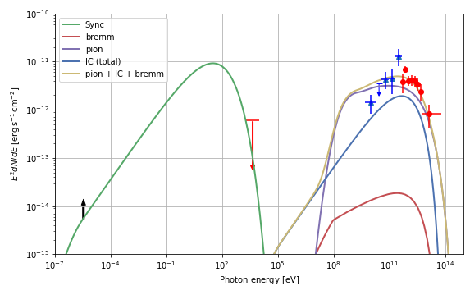}
    \caption{Data points are same as in Fig. \ref{fig:SED1}, here compared to the hadronic model described in the text.}
    \label{fig:SED2}
\end{figure}

With the robust X-ray upper limit of the entire SNR obtained in this work, we investigate whether the relativistic particle distribution in the SNR can be further constrained through its expected emission. Radio data at 843 MHz come from MGPS, the
flux point needs to be treated as a lower limit since certain angular scales may be missed. Sources with diameter above 25 arcmin are not entirely covered by the radio telescope \citep{green2014second}. The GeV spectrum obtained with
Fermi-LAT is taken from \cite{araya2017detection}. TeV data from H.E.S.S. are taken from \cite{2018A&A...612A...8H}. To convert the integral upper limit obtained from the X-ray data into a value suitable for the (differential) SEDs shown in Fig. \ref{fig:SED1} and Fig. \ref{fig:SED2}, we perform a spectral joint fitting of \textit{XMM-Newton} and \textit{Suzaku} in the range 2.0-7.0\,keV and extrapolated to 2.0-10.0\,keV (Section \ref{sec:Results1}). The \textit{naima}\footnote{\url{https://naima.readthedocs.io/en/latest/api-mcmc.html}} \citep{2015ICRC...34..922Z} package is used to simulate the non-thermal radiation from assumed relativistic particle populations. All models presented here are static one-zone models, in which homogeneous distributions of particles as well as of target fields relevant for the emission are maintained.\\

In Fig. \ref{fig:SED1}, the result from a leptonic model is shown, in which the entire emission across all wavebands is dominated by leptonic emission, electron synchrotron at radio to X-ray frequencies and Inverse Compton emission in the GeV to TeV band. The photon fields that are up-scattered by the relativistic electrons are the cosmic microwave background (CMB), near infrared stellar emission (NIR) and far infrared dust (FIR), which can be described by a blackbody distribution. For the NIR field, 3000 K and 0.4 $\mathrm{eV\ cm}^{-3}$ are adopted (\cite{shibata2010possible}, \cite{2016PhRvD..94f3009V}), for the FIR field, we use 20 K and 0.8 $\mathrm{eV\ cm}^{-3}$ \citep{araya2017detection}. The distance to the SNR is set at 3.5 kpc, adopting the association with the Galactic arm Scutum-Crux \citep{2018MNRAS.480..134M}. Since an unbroken power law is adopted for the particle population, the adopted magnetic field needs to be lower than 10 µG in order not to undershoot the radio lower limit and overshoot the $\gamma$-ray data. $\sim$10 µG would be required in simple shock-compression scenarios, and is also a value found e.g. by \cite{ferriere2009interstellar} even for the diffuse intercloud medium. Here, we adopt 6 µG \citep{araya2017detection}. To match the $\gamma$-ray spectral shape, a particle cut-off energy at 7 TeV and a spectral index of 1.9 is chosen. The gas density is assumed to be low so that Bremsstrahlung can be ignored; usually, it is assumed that if Bremsstrahlung plays a significant role in the $\gamma$-ray range, corresponding line emission in the X-ray band should be visible.  This model satisfies all data, including the upper limit derived in X-rays.  \\

Triggered by the low required magnetic field in the leptonic model which speaks against recent TeV particle acceleration,
and by the Fe K$\alpha$-indications of lower energy protons in the SNR, we also attempted to fit a hadronic emission model to the broadband data. Here, the leptonic population is only constrained by the radio to X-ray band, except for the magnetic field. In this case, we used 12 $\mu$G in order to follow the radio lower limit and at the same time reduce the contribution of IC emission in the very high energy range. Canonically, the proton particle distribution is described with a power law of energy index 2.0 (in the test-particle approach of a Fermi-accelerated spectrum), or softer if e.g. particle escape is relevant. We show a model with an index of 2.0 and a cut-off energy at 60 TeV in Fig. \ref{fig:SED2}. The input for the synchrotron and IC radiation are the same as for the leptonic model. The hadronic model requires the energy of the accelerated protons to be no more then a few  $ 10^{50}\mathrm{erg}$ \citep{1999A&A...351..330A}. 
We assume two different values of the ISM density at 0.5 $\mathrm{cm}^{-3}$ and 1 $\mathrm{cm}^{-3}$ (Fig. \ref{fig:SED2}), which corresponds to the total energy of the protons at $3.3 \times 10^{50}\mathrm{erg}$ and $1.6 \times 10^{50}\mathrm{erg}$, respectively. Such densities could be viable in a scenario where the emission happens in molecular clouds or in gas shells created by the progenitor wind. At this level of target gas density, the contribution of Bremsstrahlung emission is negligible.\\

\section{Discussion}
\label{sec:Discussion}
A pure leptonic model can well describe all available data from HESS J1534-571. The model is very similar to the one proposed by \cite{araya2017detection}. Given the relatively low energy of the particle spectral cut-off and the large size of the remnant which leads to a by orders of magnitude worse sensitivity in the X-ray band compared to typical point-source sensitivities, the X-ray limit obtained in this work is not sufficient to challenge a pure leptonic model. The low deduced magnetic field implies that the bulk of emission occurs outside of shock-acceleration regions. This is reminiscent of “relic” TeV pulsar wind nebulae, which also have no morphologically matching counterpart in X-rays \citep{2012ASPC..466..167K,2018A&A...612A...2H}. The fact that no X-ray synchrotron emission is detected even from the actual accelerating sites (possibly at the forward shock as outlined by the radio remnant) can be explained by meanwhile slowed-down shocks which imply that the corresponding high-energy end of the synchrotron spectrum has shifted to lower energies to which X-ray satellite observations are not sensitive \citep{2012A&ARv..20...49V}. Alternatively, the flux level at these sites could be too low for current detectors, given the large angular extension of the source.\\

The alternative, hadronic scenario for the $\gamma$-ray emission matches - in the simple test-particle and one-zone form as presented in Fig. \ref{fig:SED2} - the available $\gamma$-ray data less well, and might require modifications. In fact, the $\gamma$-ray spectral shape is reminiscent of the one of RX J1713.7-3946 \citep{2000A&A...354L..57M}. For this SNR, motivated by the excellent morphological match between TeV emission and gas tracers (Sect. \ref{sec:Results2}), a scenario has been proposed in which the energy-dependent penetration depth of CR protons into dense, clumpy molecular clouds leads to a strong modification of the spatially-integrated $\gamma$-ray spectrum \citep{2006A&A...449..223A,2014MNRAS.445L..70G}. Since the highest energy protons “see” denser material, the emission is boosted, which leads to a steepening of the spectrum below the cutoff energy. Such a spectrum would qualitatively better match the observed $\gamma$-ray data from HESS J1534-571. Such a scenario would imply that the MeV/GeV emission would spatially differ from the TeV emission, which is not possible to probe with the currently available $\gamma$-ray data from Fermi-LAT. However, if the Fe K$\alpha$ emission tentatively detected from HESS J1534-571 is taken as a tracer of low-energy CRs, then the apparent morphology seen in Fig. \ref{fig:6365-image-with-suzaku} is in qualitative agreement with expectations. The Fe K$\alpha$-emission is on the scale of the SNR size following the TeV emission, but on smaller angular scales there seems to be no good agreement. Moreover, such a hadronic scenario would imply the existence of dense molecular clumps in spatial correspondence to the TeV emission, and indeed, Maxted et al. (2018) have identified such molecular clumps in CO emission, following the TeV emission morphology on the scale of the full SNR. However, given the low statistical significance of individual morphological features both in the TeV and the X-ray band, any firm conclusion needs to be corrobated by more sensitive, future observations. \\

\begin{figure}
    \centering
    \includegraphics[width=3.5in]{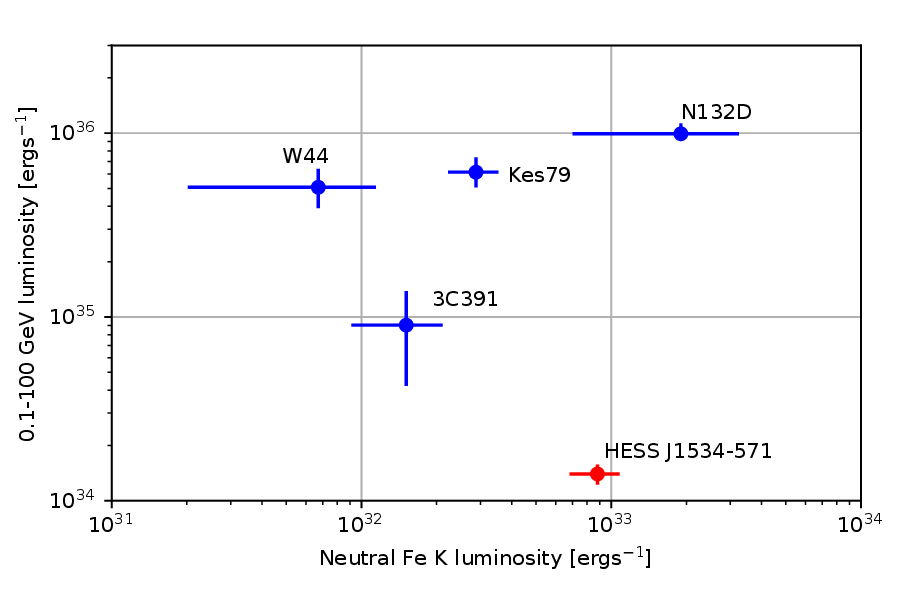}
    \caption{The relation of the Fe K$\alpha$ line emission luminosity and the luminosity in the range of 0.1-100 GeV $\gamma$-ray detection. The red data point is for HESS J1534-571 (Fe K$\alpha$: this work, GeV: \cite{araya2017detection}), blue points represent N132D, Kes79, W44, 3C391 \citep{bamba2018transition,2016ApJS..224....8A,2016PASJ...68S...8S}. }
    \label{fig:aya-san}
\end{figure}

If the Fe K$\alpha$ line at 6.4\,keV indeed stems from CR protons interacting with gas material (at typical particle energies
of 100 MeV), a correlation between the GeV luminosity and the 6.4\,keV line luminosity would be expected, not only on smaller morphological scales in an individual SNR, but also on average for all SNRs for which the $\gamma$-ray emission is suspected to be of hadronic nature. Fig. \ref{fig:aya-san} shows such a correlation plot, where we compare the data from HESS J1534-571 to numbers from other SNRs (taken from \citet{bamba2018transition}). 
To derive the line luminosity for HESS J1534-571, given that there is minimal overlap between the corresponding extraction areas, we simply added the line fluxes from \textit{XMM-Newton} ($1.9 \pm 1.0 \times  10^{-5}\ \mathrm{photons/cm^2/s}$, see Table \ref{tab:6.4table}) and from \textit{Suzaku} ($4.0 \pm 0.9 \times  10^{-5}\ \mathrm{photons/cm^2/s}$ derived from the enhanced regions in our analysis).
The statistical uncertainties of the Fe line luminosities are calculated from the respective photon flux errors at 90\% confidence level \citep{bamba2018transition, 2014ApJ...783...32A}. 
The statistical uncertainties of the GeV luminosities 
are derived from the Fermi-LAT analysis of each SNR (W44, \citet{2022ApJS..260...53A}; Kes 79, \citet{2014ApJ...783...32A}; N132D, \citet{bamba2018transition}; 3C391, \citet{2010ApJ...717..372C}; HESS J1534-571, \citet{araya2017detection}). We again adopt the distance of 3.5\,kpc following the Scutum-Crux Arm gas association in \cite{2018MNRAS.480..134M} for HESS J1534-571. As can be seen from the figure, however, no clear conclusion can be drawn for HESS J1534-571 from this comparison at this point in time. While the Fe K$\alpha$ luminosity is in the same ballpark as the one from the other SNRs, the GeV luminosity is substantially lower. This lack of dense molecular clouds in the environment may limit the intensity of the 6.4\,keV line. If confirmed, this would indicate a different physical mechanism for the production of the Fe K$\alpha$ line emission.\\

\section{Conclusion}
\label{sec:Conclusion}
For the first time, a TeV-emitting SNR has been shown to not exhibit a clear X-ray continuum counterpart at current X-ray satellite sensitivity, which is a quite robust statement given that the \textit{XMM-Newton} and \textit{Suzaku} observations together cover $\sim$85\% of the radio shell.

The $\gamma$-ray data of HESS J1534-571 can be explained in a leptonic emission scenario, with a magnetic field that indicates that the $\gamma$-ray emitting electrons have propagated away from their acceleration sites, similar to “relic” TeV pulsar wind nebula scenarios. A hadronic emission scenario can be constructed under the assumption that the spectrum is modified by energy-dependent penetration of the accelerated protons into dense and clumpy gas material. Such a scenario also avoids a potential conflict with the lack of thermal X-ray emission from the SNR. We find evidence for Fe K$\alpha$ emission from localized regions in the SNR, confirming earlier findings of \cite{saji2018discovery}. Whether this emission is caused by interactions of low-energy CRs in the SNR and thus could support a hadronic nature of the $\gamma$-ray emission requires further investigations with more sensitive, next generation instruments.

\section*{Acknowledgement}
\addcontentsline{toc}{section}{Acknowledgements}
We acknowledge support from the Deutsches Zentrum f\"ur Luft- und Raumfahrt (DLR) through DLR-PT grant FKZ 50 OR 1914. This work is based on observations obtained with \textit{XMM-Newton}, an ESA science mission with instruments and contributions directly funded by ESA Member States and NASA. This research has also made use of data obtained from the \textit{Suzaku} satellite, a collaborative mission between the space agencies of Japan (JAXA) and the USA (NASA). 

\bibliographystyle{aa}
\bibliography{references}

\begin{thebibliography}{46}
\expandafter\ifx\csname natexlab\endcsname\relax\def\natexlab#1{#1}\fi

\bibitem[{{Abdollahi} {et~al.}(2022){Abdollahi}, {Acero}, {Baldini}, {Ballet},
  {Bastieri}, {Bellazzini}, {Berenji}, {Berretta}, {Bissaldi}, {Blandford},
  {Bloom}, {Bonino}, {Brill}, {Britto}, {Bruel}, {Burnett}, {Buson}, {Cameron},
  {Caputo}, {Caraveo}, {Castro}, {Chaty}, {Cheung}, {Chiaro}, {Cibrario},
  {Ciprini}, {Coronado-Bl{\'a}zquez}, {Crnogorcevic}, {Cutini}, {D'Ammando},
  {De Gaetano}, {Digel}, {Di Lalla}, {Dirirsa}, {Di Venere}, {Dom{\'\i}nguez},
  {Fallah Ramazani}, {Fegan}, {Ferrara}, {Fiori}, {Fleischhack}, {Franckowiak},
  {Fukazawa}, {Funk}, {Fusco}, {Galanti}, {Gammaldi}, {Gargano}, {Garrappa},
  {Gasparrini}, {Giacchino}, {Giglietto}, {Giordano}, {Giroletti}, {Glanzman},
  {Green}, {Grenier}, {Grondin}, {Guillemot}, {Guiriec}, {Gustafsson},
  {Harding}, {Hays}, {Hewitt}, {Horan}, {Hou}, {J{\'o}hannesson}, {Karwin},
  {Kayanoki}, {Kerr}, {Kuss}, {Landriu}, {Larsson}, {Latronico},
  {Lemoine-Goumard}, {Li}, {Liodakis}, {Longo}, {Loparco}, {Lott}, {Lubrano},
  {Maldera}, {Malyshev}, {Manfreda}, {Mart{\'\i}-Devesa}, {Mazziotta}, {Mereu},
  {Meyer}, {Michelson}, {Mirabal}, {Mitthumsiri}, {Mizuno}, {Moiseev},
  {Monzani}, {Morselli}, {Moskalenko}, {Negro}, {Nuss}, {Omodei}, {Orienti},
  {Orlando}, {Paneque}, {Pei}, {Perkins}, {Persic}, {Pesce-Rollins},
  {Petrosian}, {Pillera}, {Poon}, {Porter}, {Principe}, {Rain{\`o}}, {Rando},
  {Rani}, {Razzano}, {Razzaque}, {Reimer}, {Reimer}, {Reposeur},
  {S{\'a}nchez-Conde}, {Saz Parkinson}, {Scotton}, {Serini}, {Sgr{\`o}},
  {Siskind}, {Smith}, {Spandre}, {Spinelli}, {Sueoka}, {Suson}, {Tajima},
  {Tak}, {Thayer}, {Thompson}, {Torres}, {Troja}, {Valverde}, {Wood}, \&
  {Zaharijas}}]{2022ApJS..260...53A}
{Abdollahi}, S., {Acero}, F., {Baldini}, L., {et~al.} 2022, \apjs, 260, 53

\bibitem[{{Acero} {et~al.}(2016){Acero}, {Ackermann}, {Ajello}, {Baldini},
  {Ballet}, {Barbiellini}, {Bastieri}, {Bellazzini}, {Bissaldi}, {Blandford},
  {Bloom}, {Bonino}, {Bottacini}, {Brandt}, {Bregeon}, {Bruel}, {Buehler},
  {Buson}, {Caliandro}, {Cameron}, {Caputo}, {Caragiulo}, {Caraveo},
  {Casandjian}, {Cavazzuti}, {Cecchi}, {Chekhtman}, {Chiang}, {Chiaro},
  {Ciprini}, {Claus}, {Cohen}, {Cohen-Tanugi}, {Cominsky}, {Condon}, {Conrad},
  {Cutini}, {D'Ammando}, {de Angelis}, {de Palma}, {Desiante}, {Digel}, {Di
  Venere}, {Drell}, {Drlica-Wagner}, {Favuzzi}, {Ferrara}, {Franckowiak},
  {Fukazawa}, {Funk}, {Fusco}, {Gargano}, {Gasparrini}, {Giglietto}, {Giommi},
  {Giordano}, {Giroletti}, {Glanzman}, {Godfrey}, {Gomez-Vargas}, {Grenier},
  {Grondin}, {Guillemot}, {Guiriec}, {Gustafsson}, {Hadasch}, {Harding},
  {Hayashida}, {Hays}, {Hewitt}, {Hill}, {Horan}, {Hou}, {Iafrate}, {Jogler},
  {J{\'o}hannesson}, {Johnson}, {Kamae}, {Katagiri}, {Kataoka}, {Katsuta},
  {Kerr}, {Kn{\"o}dlseder}, {Kocevski}, {Kuss}, {Laffon}, {Lande}, {Larsson},
  {Latronico}, {Lemoine-Goumard}, {Li}, {Li}, {Longo}, {Loparco}, {Lovellette},
  {Lubrano}, {Magill}, {Maldera}, {Marelli}, {Mayer}, {Mazziotta}, {Michelson},
  {Mitthumsiri}, {Mizuno}, {Moiseev}, {Monzani}, {Moretti}, {Morselli},
  {Moskalenko}, {Murgia}, {Nemmen}, {Nuss}, {Ohsugi}, {Omodei}, {Orienti},
  {Orlando}, {Ormes}, {Paneque}, {Perkins}, {Pesce-Rollins}, {Petrosian},
  {Piron}, {Pivato}, {Porter}, {Rain{\`o}}, {Rando}, {Razzano}, {Razzaque},
  {Reimer}, {Reimer}, {Renaud}, {Reposeur}, {Rousseau}, {Saz Parkinson},
  {Schmid}, {Schulz}, {Sgr{\`o}}, {Siskind}, {Spada}, {Spandre}, {Spinelli},
  {Strong}, {Suson}, {Tajima}, {Takahashi}, {Tanaka}, {Thayer}, {Thompson},
  {Tibaldo}, {Tibolla}, {Torres}, {Tosti}, {Troja}, {Uchiyama}, {Vianello},
  {Wells}, {Wood}, {Wood}, {Yassine}, {den Hartog}, \&
  {Zimmer}}]{2016ApJS..224....8A}
{Acero}, F., {Ackermann}, M., {Ajello}, M., {et~al.} 2016, \apjs, 224, 8

\bibitem[{{Aharonian} {et~al.}(2006){Aharonian}, {Akhperjanian}, {Bazer-Bachi},
  {Beilicke}, {Benbow}, {Berge}, {Bernl{\"o}hr}, {Boisson}, {Bolz}, {Borrel},
  {Braun}, {Breitling}, {Brown}, {Chadwick}, {Chounet}, {Cornils},
  {Costamante}, {Degrange}, {Dickinson}, {Djannati-Ata{\"\i}}, {O'C. Drury},
  {Dubus}, {Emmanoulopoulos}, {Espigat}, {Feinstein}, {Fontaine}, {Fuchs},
  {Funk}, {Gallant}, {Giebels}, {Glicenstein}, {Goret}, {Hadjichristidis},
  {Hauser}, {Hauser}, {Heinzelmann}, {Henri}, {Hermann}, {Hinton}, {Hofmann},
  {Holleran}, {Horns}, {Jacholkowska}, {de Jager}, {Kh{\'e}lifi}, {Klages},
  {Komin}, {Konopelko}, {Latham}, {Le Gallou}, {Lemi{\`e}re},
  {Lemoine-Goumard}, {Lohse}, {Martin}, {Martineau-Huynh}, {Marcowith},
  {Masterson}, {McComb}, {de Naurois}, {Nedbal}, {Nolan}, {Noutsos}, {Orford},
  {Osborne}, {Ouchrif}, {Panter}, {Pelletier}, {Pita}, {P{\"u}hlhofer},
  {Punch}, {Raubenheimer}, {Raue}, {Rayner}, {Reimer}, {Reimer}, {Ripken},
  {Rob}, {Rolland}, {Rowell}, {Sahakian}, {Saug{\'e}}, {Schlenker},
  {Schlickeiser}, {Schuster}, {Schwanke}, {Siewert}, {Sol}, {Spangler},
  {Steenkamp}, {Stegmann}, {Superina}, {Tavernet}, {Terrier}, {Th{\'e}oret},
  {Tluczykont}, {van Eldik}, {Vasileiadis}, {Venter}, {Vincent}, {V{\"o}lk}, \&
  {Wagner}}]{2006A&A...449..223A}
{Aharonian}, F., {Akhperjanian}, A.~G., {Bazer-Bachi}, A.~R., {et~al.} 2006,
  \aap, 449, 223

\bibitem[{{Aharonian} \& {Atoyan}(1999)}]{1999A&A...351..330A}
{Aharonian}, F.~A. \& {Atoyan}, A.~M. 1999, \aap, 351, 330

\bibitem[{Araya(2017)}]{araya2017detection}
Araya, M. 2017, The Astrophysical Journal, 843, 12

\bibitem[{{Auchettl} {et~al.}(2014){Auchettl}, {Slane}, \&
  {Castro}}]{2014ApJ...783...32A}
{Auchettl}, K., {Slane}, P., \& {Castro}, D. 2014, \apj, 783, 32

\bibitem[{Bamba {et~al.}(2018)Bamba, Ohira, Yamazaki, Sawada, Terada, Koyama,
  Miller, Yamaguchi, Katsuda, Nobukawa, {et~al.}}]{bamba2018transition}
Bamba, A., Ohira, Y., Yamazaki, R., {et~al.} 2018, The Astrophysical Journal,
  854, 71

\bibitem[{{Bell}(2004)}]{2004MNRAS.353..550B}
{Bell}, A.~R. 2004, \mnras, 353, 550

\bibitem[{{Bell} \& {Lucek}(2001)}]{2001MNRAS.321..433B}
{Bell}, A.~R. \& {Lucek}, S.~G. 2001, \mnras, 321, 433

\bibitem[{{Blandford} \& {Eichler}(1987)}]{1987PhR...154....1B}
{Blandford}, R. \& {Eichler}, D. 1987, \physrep, 154, 1

\bibitem[{{Burton} {et~al.}(2013){Burton}, {Braiding}, {Glueck}, {Goldsmith},
  {Hawkes}, {Hollenbach}, {Kulesa}, {Martin}, {Pineda}, {Rowell}, {Simon},
  {Stark}, {Stutzki}, {Tothill}, {Urquhart}, {Walker}, {Walsh}, \&
  {Wolfire}}]{2013PASA...30...44B}
{Burton}, M.~G., {Braiding}, C., {Glueck}, C., {et~al.} 2013, \pasa, 30, e044

\bibitem[{{Castro} \& {Slane}(2010)}]{2010ApJ...717..372C}
{Castro}, D. \& {Slane}, P. 2010, \apj, 717, 372

\bibitem[{{Drury} {et~al.}(1994){Drury}, {Aharonian}, \&
  {Voelk}}]{1994A&A...287..959D}
{Drury}, L.~O., {Aharonian}, F.~A., \& {Voelk}, H.~J. 1994, \aap, 287, 959

\bibitem[{Ferriere(2009)}]{ferriere2009interstellar}
Ferriere, K. 2009, Astronomy \& Astrophysics, 505, 1183

\bibitem[{{Gabici} \& {Aharonian}(2014)}]{2014MNRAS.445L..70G}
{Gabici}, S. \& {Aharonian}, F.~A. 2014, \mnras, 445, L70

\bibitem[{{Ginzburg} \& {Syrovatskii}(1964)}]{1964ocr..book.....G}
{Ginzburg}, V.~L. \& {Syrovatskii}, S.~I. 1964, {The Origin of Cosmic Rays}

\bibitem[{Green {et~al.}(2014)Green, Reeves, \& Murphy}]{green2014second}
Green, A., Reeves, S., \& Murphy, T. 2014, Publications of the Astronomical
  Society of Australia, 31

\bibitem[{{H.~E.~S.~S. Collaboration} {et~al.}(2018{\natexlab{a}}){H.~E.~S.~S.
  Collaboration}, {Abdalla}, {Abramowski}, {Aharonian}, {Ait Benkhali},
  {Akhperjanian}, {Andersson}, {Ang{\"u}ner}, {Arakawa}, {Arrieta}, {Aubert},
  {Backes}, {Balzer}, {Barnard}, {Becherini}, {Becker Tjus}, {Berge},
  {Bernhard}, {Bernl{\"o}hr}, {Blackwell}, {B{\"o}ttcher}, {Boisson},
  {Bolmont}, {Bonnefoy}, {Bordas}, {Bregeon}, {Brun}, {Brun}, {Bryan},
  {B{\"u}chele}, {Bulik}, {Capasso}, {Carr}, {Casanova}, {Cerruti},
  {Chakraborty}, {Chaves}, {Chen}, {Chevalier}, {Coffaro}, {Colafrancesco},
  {Cologna}, {Condon}, {Conrad}, {Cui}, {Davids}, {Decock}, {Degrange}, {Deil},
  {Devin}, {deWilt}, {Dirson}, {Djannati-Ata{\"\i}}, {Domainko}, {Donath},
  {Drury}, {Dutson}, {Dyks}, {Edwards}, {Egberts}, {Eger}, {Ernenwein},
  {Eschbach}, {Farnier}, {Fegan}, {Fernandes}, {Fiasson}, {Fontaine},
  {F{\"o}rster}, {Funk}, {F{\"u}{\ss}ling}, {Gabici}, {Gajdus}, {Gallant},
  {Garrigoux}, {Giavitto}, {Giebels}, {Glicenstein}, {Gottschall}, {Goyal},
  {Grondin}, {Hahn}, {Haupt}, {Hawkes}, {Heinzelmann}, {Henri}, {Hermann},
  {Hervet}, {Hinton}, {Hofmann}, {Hoischen}, {Holch}, {Holler}, {Horns},
  {Ivascenko}, {Iwasaki}, {Jacholkowska}, {Jamrozy}, {Janiak}, {Jankowsky},
  {Jankowsky}, {Jingo}, {Jogler}, {Jouvin}, {Jung-Richardt}, {Kastendieck},
  {Katarzy{\'n}ski}, {Katsuragawa}, {Katz}, {Kerszberg}, {Khangulyan},
  {Kh{\'e}lifi}, {King}, {Klepser}, {Klochkov}, {Klu{\'z}niak}, {Kolitzus},
  {Komin}, {Kosack}, {Krakau}, {Kraus}, {Kr{\"u}ger}, {Laffon}, {Lamanna},
  {Lau}, {Lees}, {Lefaucheur}, {Lefranc}, {Lemi{\`e}re}, {Lemoine-Goumard},
  {Lenain}, {Leser}, {Lohse}, {Lorentz}, {Liu}, {L{\'o}pez-Coto}, {Lypova},
  {Marandon}, {Marcowith}, {Mariaud}, {Marx}, {Maurin}, {Maxted}, {Mayer},
  {Meintjes}, {Meyer}, {Mitchell}, {Moderski}, {Mohamed}, {Mohrmann},
  {Mor{\r{a}}}, {Moulin}, {Murach}, {Nakashima}, {de Naurois}, {Niederwanger},
  {Niemiec}, {Oakes}, {O'Brien}, {Odaka}, {{\"O}ttl}, {Ohm}, {Ostrowski},
  {Oya}, {Padovani}, {Panter}, {Parsons}, {Pekeur}, {Pelletier}, {Perennes},
  {Petrucci}, {Peyaud}, {Piel}, {Pita}, {Poon}, {Prokhorov}, {Prokoph},
  {P{\"u}hlhofer}, {Punch}, {Quirrenbach}, {Raab}, {Reimer}, {Reimer},
  {Renaud}, {de los Reyes}, {Richter}, {Rieger}, {Romoli}, {Rowell}, {Rudak},
  {Rulten}, {Sahakian}, {Saito}, {Salek}, {Sanchez}, {Santangelo}, {Sasaki},
  {Schlickeiser}, {Sch{\"u}ssler}, {Schulz}, {Schwanke}, {Schwemmer},
  {Seglar-Arroyo}, {Settimo}, {Seyffert}, {Shafi}, {Shilon}, {Simoni}, {Sol},
  {Spanier}, {Spengler}, {Spies}, {Stawarz}, {Steenkamp}, {Stegmann}, {Stycz},
  {Sushch}, {Takahashi}, {Tavernet}, {Tavernier}, {Taylor}, {Terrier},
  {Tibaldo}, {Tiziani}, {Tluczykont}, {Trichard}, {Tsuji}, {Tuffs}, {Uchiyama},
  {van der Walt}, {van Eldik}, {van Rensburg}, {van Soelen}, {Vasileiadis},
  {Veh}, {Venter}, {Viana}, {Vincent}, {Vink}, {Voisin}, {V{\"o}lk},
  {Vuillaume}, {Wadiasingh}, {Wagner}, {Wagner}, {Wagner}, {White},
  {Wierzcholska}, {Willmann}, {W{\"o}rnlein}, {Wouters}, {Yang}, {Zabalza},
  {Zaborov}, {Zacharias}, {Zanin}, {Zdziarski}, {Zech}, {Zefi}, {Ziegler},
  {{\.Z}ywucka}, {Bamba}, {Fukui}, {Sano}, \& {Yoshiike}}]{2018A&A...612A...8H}
{H.~E.~S.~S. Collaboration}, {Abdalla}, H., {Abramowski}, A., {et~al.}
  2018{\natexlab{a}}, \aap, 612, A8

\bibitem[{{H.~E.~S.~S. Collaboration} {et~al.}(2018{\natexlab{b}}){H.~E.~S.~S.
  Collaboration}, {Abdalla}, {Abramowski}, {Aharonian}, {Ait Benkhali},
  {Akhperjanian}, {Andersson}, {Ang{\"u}ner}, {Arrieta}, {Aubert}, {Backes},
  {Balzer}, {Barnard}, {Becherini}, {Becker Tjus}, {Berge}, {Bernhard},
  {Bernl{\"o}hr}, {Blackwell}, {B{\"o}ttcher}, {Boisson}, {Bolmont}, {Bordas},
  {Bregeon}, {Brun}, {Brun}, {Bryan}, {Bulik}, {Capasso}, {Carr}, {Carrigan},
  {Casanova}, {Cerruti}, {Chakraborty}, {Chalme-Calvet}, {Chaves}, {Chen},
  {Chevalier}, {Chr{\'e}tien}, {Colafrancesco}, {Cologna}, {Condon}, {Conrad},
  {Couturier}, {Cui}, {Davids}, {Degrange}, {Deil}, {Devin}, {deWilt},
  {Dirson}, {Djannati-Ata{\"\i}}, {Domainko}, {Donath}, {Drury}, {Dubus},
  {Dutson}, {Dyks}, {Edwards}, {Egberts}, {Eger}, {Ernenwein}, {Eschbach},
  {Farnier}, {Fegan}, {Fernandes}, {Fiasson}, {Fontaine}, {F{\"o}rster},
  {Funk}, {F{\"u}{\ss}ling}, {Gabici}, {Gajdus}, {Gallant}, {Garrigoux},
  {Giavitto}, {Giebels}, {Glicenstein}, {Gottschall}, {Goyal}, {Grondin},
  {Hadasch}, {Hahn}, {Haupt}, {Hawkes}, {Heinzelmann}, {Henri}, {Hermann},
  {Hervet}, {Hillert}, {Hinton}, {Hofmann}, {Hoischen}, {Holler}, {Horns},
  {Ivascenko}, {Jacholkowska}, {Jamrozy}, {Janiak}, {Jankowsky}, {Jankowsky},
  {Jingo}, {Jogler}, {Jouvin}, {Jung-Richardt}, {Kastendieck},
  {Katarzy{\'n}ski}, {Katz}, {Kerszberg}, {Kh{\'e}lifi}, {Kieffer}, {King},
  {Klepser}, {Klochkov}, {Klu{\'z}niak}, {Kolitzus}, {Komin}, {Kosack},
  {Krakau}, {Kraus}, {Krayzel}, {Kr{\"u}ger}, {Laffon}, {Lamanna}, {Lau},
  {Lees}, {Lefaucheur}, {Lefranc}, {Lemi{\`e}re}, {Lemoine-Goumard}, {Lenain},
  {Leser}, {Lohse}, {Lorentz}, {Liu}, {L{\'o}pez-Coto}, {Lypova}, {Marandon},
  {Marcowith}, {Mariaud}, {Marx}, {Maurin}, {Maxted}, {Mayer}, {Meintjes},
  {Meyer}, {Mitchell}, {Moderski}, {Mohamed}, {Mohrmann}, {Mor{\r{a}}},
  {Moulin}, {Murach}, {de Naurois}, {Niederwanger}, {Niemiec}, {Oakes},
  {O'Brien}, {Odaka}, {{\"O}ttl}, {Ohm}, {de O{\~n}a Wilhelmi}, {Ostrowski},
  {Oya}, {Padovani}, {Panter}, {Parsons}, {Paz Arribas}, {Pekeur}, {Pelletier},
  {Perennes}, {Petrucci}, {Peyaud}, {Pita}, {Poon}, {Prokhorov}, {Prokoph},
  {P{\"u}hlhofer}, {Punch}, {Quirrenbach}, {Raab}, {Reimer}, {Reimer},
  {Renaud}, {de los Reyes}, {Rieger}, {Romoli}, {Rosier-Lees}, {Rowell},
  {Rudak}, {Rulten}, {Sahakian}, {Salek}, {Sanchez}, {Santangelo}, {Sasaki},
  {Schlickeiser}, {Sch{\"u}ssler}, {Schulz}, {Schwanke}, {Schwemmer},
  {Settimo}, {Seyffert}, {Shafi}, {Shilon}, {Simoni}, {Sol}, {Spanier},
  {Spengler}, {Spies}, {Stawarz}, {Steenkamp}, {Stegmann}, {Stinzing}, {Stycz},
  {Sushch}, {Tavernet}, {Tavernier}, {Taylor}, {Terrier}, {Tibaldo}, {Tiziani},
  {Tluczykont}, {Trichard}, {Tuffs}, {Uchiyama}, {Valerius}, {van der Walt},
  {van Eldik}, {van Soelen}, {Vasileiadis}, {Veh}, {Venter}, {Viana},
  {Vincent}, {Vink}, {Voisin}, {V{\"o}lk}, {Vuillaume}, {Wadiasingh}, {Wagner},
  {Wagner}, {Wagner}, {White}, {Wierzcholska}, {Willmann}, {W{\"o}rnlein},
  {Wouters}, {Yang}, {Zabalza}, {Zaborov}, {Zacharias}, {Zdziarski}, {Zech},
  {Zefi}, {Ziegler}, \& {{\.Z}ywucka}}]{2018A&A...612A...2H}
{H.~E.~S.~S. Collaboration}, {Abdalla}, H., {Abramowski}, A., {et~al.}
  2018{\natexlab{b}}, \aap, 612, A2

\bibitem[{{HI4PI Collaboration} {et~al.}(2016){HI4PI Collaboration}, {Ben
  Bekhti}, {Fl{\"o}er}, {Keller}, {Kerp}, {Lenz}, {Winkel}, {Bailin},
  {Calabretta}, {Dedes}, {Ford}, {Gibson}, {Haud}, {Janowiecki}, {Kalberla},
  {Lockman}, {McClure-Griffiths}, {Murphy}, {Nakanishi}, {Pisano}, \&
  {Staveley-Smith}}]{2016A&A...594A.116H}
{HI4PI Collaboration}, {Ben Bekhti}, N., {Fl{\"o}er}, L., {et~al.} 2016, \aap,
  594, A116

\bibitem[{{Hillas}(2005)}]{2005JPhG...31R..95H}
{Hillas}, A.~M. 2005, Journal of Physics G Nuclear Physics, 31, R95

\bibitem[{{Jones} \& {Ellison}(1991)}]{1991SSRv...58..259J}
{Jones}, F.~C. \& {Ellison}, D.~C. 1991, \ssr, 58, 259

\bibitem[{{Kaastra} \& {Mewe}(1993)}]{1993A&AS...97..443K}
{Kaastra}, J.~S. \& {Mewe}, R. 1993, \aaps, 97, 443

\bibitem[{Kallman {et~al.}(2004)Kallman, Palmeri, Bautista, Mendoza, \&
  Krolik}]{kallman2004photoionization}
Kallman, T., Palmeri, P., Bautista, M., Mendoza, C., \& Krolik, J. 2004, The
  Astrophysical Journal Supplement Series, 155, 675

\bibitem[{{Kargaltsev} {et~al.}(2012){Kargaltsev}, {Pavlov}, \&
  {Durant}}]{2012ASPC..466..167K}
{Kargaltsev}, O., {Pavlov}, G.~G., \& {Durant}, M. 2012, in Astronomical
  Society of the Pacific Conference Series, Vol. 466, Electromagnetic Radiation
  from Pulsars and Magnetars, ed. W.~{Lewandowski}, O.~{Maron}, \& J.~{Kijak},
  167

\bibitem[{{Kostrzewa-Rutkowska} {et~al.}(2017){Kostrzewa-Rutkowska}, {Lopez},
  {Jonker}, {Chen}, {Fraser}, {Sollerman}, {Inserra}, {Kankare}, {Maguire},
  {Smartt}, {Smith}, {Sullivan}, {Valenti}, {Yaron}, {Young}, {Manulis},
  {Tonry}, {Stalder}, {Denneau}, {Heinze}, {Weiland}, {Rest}, {Chambers},
  {Flewelling}, {Huber}, {Magnier}, {Schultz}, {Waters}, {Wainscoat}, \&
  {Wilman}}]{2017ATel11007....1K}
{Kostrzewa-Rutkowska}, Z., {Lopez}, K.~M., {Jonker}, P.~G., {et~al.} 2017, The
  Astronomer's Telegram, 11007, 1

\bibitem[{{Koyama} {et~al.}(2007){Koyama}, {Hyodo}, {Inui}, {Nakajima},
  {Matsumoto}, {Tsuru}, {Takahashi}, {Maeda}, {Yamazaki}, {Murakami},
  {Yamauchi}, {Tsuboi}, {Senda}, {Kataoka}, {Takahashi}, {Holt}, \&
  {Brown}}]{2007PASJ...59S.245K}
{Koyama}, K., {Hyodo}, Y., {Inui}, T., {et~al.} 2007, \pasj, 59, 245

\bibitem[{{Krause}(1979)}]{1979JPCRD...8..307K}
{Krause}, M.~O. 1979, Journal of Physical and Chemical Reference Data, 8, 307

\bibitem[{Kuntz \& Snowden(2008)}]{kuntz2008epic}
Kuntz, K. \& Snowden, S. 2008, Astronomy \& Astrophysics, 478, 575

\bibitem[{{Lumb} {et~al.}(2012){Lumb}, {Schartel}, \&
  {Jansen}}]{2012arXiv1202.1651L}
{Lumb}, D.~H., {Schartel}, N., \& {Jansen}, F.~A. 2012, arXiv e-prints,
  arXiv:1202.1651

\bibitem[{{Malkov} \& {Drury}(2001)}]{2001RPPh...64..429M}
{Malkov}, M.~A. \& {Drury}, L.~O. 2001, Reports on Progress in Physics, 64, 429

\bibitem[{Maxted {et~al.}(2018)Maxted, Braiding, Wong, Rowell, Burton,
  Filipovi{\'c}, Voisin, Uro{\v{s}}evi{\'c}, Vukoti{\'c}, Pavlovi{\'c},
  {et~al.}}]{maxted2018searching}
Maxted, N.~I., Braiding, C., Wong, G.~F., {et~al.} 2018, Monthly Notices of the
  Royal Astronomical Society, 480, 134

\bibitem[{{Maxted} {et~al.}(2018){Maxted}, {Braiding}, {Wong}, {Rowell},
  {Burton}, {Filipovi{\'c}}, {Voisin}, {Uro{\v{s}}evi{\'c}}, {Vukoti{\'c}},
  {Pavlovi{\'c}}, {Sano}, \& {Fukui}}]{2018MNRAS.480..134M}
{Maxted}, N.~I., {Braiding}, C., {Wong}, G.~F., {et~al.} 2018, \mnras, 480, 134

\bibitem[{{McClure-Griffiths} {et~al.}(2005){McClure-Griffiths}, {Dickey},
  {Gaensler}, {Green}, {Haverkorn}, \& {Strasser}}]{2005ApJS..158..178M}
{McClure-Griffiths}, N.~M., {Dickey}, J.~M., {Gaensler}, B.~M., {et~al.} 2005,
  \apjs, 158, 178

\bibitem[{{Muraishi} {et~al.}(2000){Muraishi}, {Tanimori}, {Yanagita},
  {Yoshida}, {Moriya}, {Kifune}, {Dazeley}, {Edwards}, {Gunji}, {Hara}, {Hara},
  {Kawachi}, {Kubo}, {Matsubara}, {Mizumoto}, {Mori}, {Muraki}, {Naito},
  {Nishijima}, {Patterson}, {Rowell}, {Sako}, {Sakurazawa}, {Susukita},
  {Tamura}, \& {Yoshikoshi}}]{2000A&A...354L..57M}
{Muraishi}, H., {Tanimori}, T., {Yanagita}, S., {et~al.} 2000, \aap, 354, L57

\bibitem[{{Negoro} {et~al.}(2017){Negoro}, {Ishikawa}, {Ueno}, {Tomida},
  {Sugawara}, {Isobe}, {Shimomukai}, {Mihara}, {Sugizaki}, {Serino}, {Iwakiri},
  {Shidatsu}, {Matsuoka}, {Kawai}, {Sugita}, {Yoshii}, {Tachibana}, {Harita},
  {Muraki}, {Morita}, {Yoshida}, {Sakamoto}, {Kawakubo}, {Kitaoka},
  {Hashimoto}, {Tsunemi}, {Yoneyama}, {Nakajima}, {Kawase}, {Sakamaki}, {Ueda},
  {Hori}, {Tanimoto}, {Oda}, {Tsuboi}, {Nakamura}, {Sasaki}, {Kawai},
  {Yamauchi}, {Hanyu}, {Hidaka}, {Kawamuro}, \&
  {Yamaoka}}]{2017ATel10699....1N}
{Negoro}, H., {Ishikawa}, M., {Ueno}, S., {et~al.} 2017, The Astronomer's
  Telegram, 10699, 1

\bibitem[{{Nobukawa} {et~al.}(2018){Nobukawa}, {Nobukawa}, {Koyama},
  {Yamauchi}, {Uchiyama}, {Okon}, {Tanaka}, {Uchida}, \&
  {Tsuru}}]{2018ApJ...854...87N}
{Nobukawa}, K.~K., {Nobukawa}, M., {Koyama}, K., {et~al.} 2018, \apj, 854, 87

\bibitem[{{Reynolds}(1998)}]{1998ApJ...493..375R}
{Reynolds}, S.~P. 1998, \apj, 493, 375

\bibitem[{Saji {et~al.}(2018)Saji, Matsumoto, Nobukawa, Nobukawa, Uchiyama,
  Yamauchi, \& Koyama}]{saji2018discovery}
Saji, S., Matsumoto, H., Nobukawa, M., {et~al.} 2018, Publications of the
  Astronomical Society of Japan, 70, 23

\bibitem[{{Sato} {et~al.}(2016){Sato}, {Koyama}, {Lee}, \&
  {Takahashi}}]{2016PASJ...68S...8S}
{Sato}, T., {Koyama}, K., {Lee}, S.-H., \& {Takahashi}, T. 2016, \pasj, 68, S8

\bibitem[{Shibata {et~al.}(2010)Shibata, Ishikawa, \&
  Sekiguchi}]{shibata2010possible}
Shibata, T., Ishikawa, T., \& Sekiguchi, S. 2010, The Astrophysical Journal,
  727, 38

\bibitem[{{Sturner} {et~al.}(1997){Sturner}, {Skibo}, {Dermer}, \&
  {Mattox}}]{1997ApJ...490..619S}
{Sturner}, S.~J., {Skibo}, J.~G., {Dermer}, C.~D., \& {Mattox}, J.~R. 1997,
  \apj, 490, 619

\bibitem[{{Vernetto} \& {Lipari}(2016)}]{2016PhRvD..94f3009V}
{Vernetto}, S. \& {Lipari}, P. 2016, \prd, 94, 063009

\bibitem[{{Vink}(2012)}]{2012A&ARv..20...49V}
{Vink}, J. 2012, \aapr, 20, 49

\bibitem[{{Yamauchi} {et~al.}(2016){Yamauchi}, {Nobukawa}, {Nobukawa},
  {Uchiyama}, \& {Koyama}}]{2016PASJ...68...59Y}
{Yamauchi}, S., {Nobukawa}, K.~K., {Nobukawa}, M., {Uchiyama}, H., \& {Koyama},
  K. 2016, \pasj, 68, 59

\bibitem[{{Zabalza}(2015)}]{2015ICRC...34..922Z}
{Zabalza}, V. 2015, in International Cosmic Ray Conference, Vol.~34, 34th
  International Cosmic Ray Conference (ICRC2015), 922

\end{thebibliography}

\end{document}